\documentclass[paper,letterpaper,notoc]{JHEP3}

\usepackage{epsfig}

\title{Markov Chain Monte Carlo Exploration of Minimal Supergravity with
Implications for Dark Matter}

\author{Edward A. Baltz\\ KIPAC\\SLAC MS 29\\2575 Sand Hill Road\\ Menlo Park,
  CA 94025\\ E-mail: \email{eabaltz@slac.stanford.edu}}

\author{Paolo Gondolo \\ Physics Department\\ University of Utah\\
115 S 1400 E, Suite 201\\ Salt Lake City, UT 84112 \\
E-mail: \email{paolo@physics.utah.edu}}


\abstract{We explore the full parameter space of Minimal Supergravity (mSUGRA),
allowing all four continuous parameters (the scalar mass $m_0$, the gaugino
mass $m_{1/2}$, the trilinear coupling $A_0$, and the ratio of Higgs vacuum
expectation values $\tan\beta$) to vary freely.  We apply current accelerator
constraints on sparticle and Higgs masses, and on the $b\rightarrow s \gamma$
branching ratio, and discuss the impact of the constraints on $g_\mu-2$\@.  To
study dark matter, we apply the WMAP constraint on the cold dark matter
density.  We develop Markov Chain Monte Carlo (MCMC) techniques to explore the
parameter regions consistent with WMAP, finding them to be considerably
superior to previously used methods for exploring supersymmetric parameter
spaces.  Finally, we study the reach of current and future direct detection
experiments in light of the WMAP constraint.}

\begin{document}

\section{Introduction}

Two of the most interesting open questions in physics, the stabilization of the
gauge hierarchy and the nature of the dark matter, naively appear to be related
in that both point to new physics at the weak scale $\sim100$ GeV\@.
Supersymmetric theories are an explicit realization of this relationship.  The
gauge hierarchy (in particular the Higgs mass) is stabilized by the interplay
between fermionic and bosonic radiative corrections.  Dark matter appears
trivially with the imposition of R-parity conservation, which stabilizes the
lightest superpartner (LSP).  The relic density of stable particles with masses
$\sim100$ GeV undergoing thermal freeze-out in the early universe is
generically found to be of the observed order.

Supersymmetric extensions to the Standard Model are plagued by a lack of
predictiveness due to the large number of undetermined parameters.  Even
assumptions that drastically reduce the parameter space leave enough parameters
that exhaustive study is difficult.  In this paper we apply the well-known (in
other fields) techniques of Markov Chain Monte Carlo (MCMC) to explore the
parameter space of Minimal Supergravity (mSUGRA), a simple and widely studied
scenario for supersymmetry breaking.  This is a four parameter model, with one
sign that must be chosen.  There are many constraints from (lack of) data on
e.g.\ the masses of the superpartners.  The most restrictive constraint comes
from the assumption that the LSP constitutes the dark matter.  The cosmological
dark matter density is known to better than 10\% accuracy.  In mSUGRA, the
parameter region consistent with the dark matter density is very thin.  Grid
methods have a very difficult time finding these models, as the volume occupied
by them in parameter space is a very small fraction of the total volume
consistent with accelerator searches.  Even with this relatively small number
of parameters we find that MCMC is much superior to grid searches.  Our future
goal is to apply MCMC techniques to larger parameter spaces, with the
expectation that only modest increases in computational resources will be
required.

\section{Minimal Supergravity}

One of the most well studied frameworks for supersymmetry breaking is Minimal
Supergravity \cite{msugra}.  This model distills the multitude of supersymmetry
breaking parameters into merely four, plus one sign.  It assumes that all
sfermions share the same mass at the GUT scale, $m_0$.  This contributes to the
masses of the Higgs doublets as well.  It assumes that the hypercharge, weak,
and strong gauginos share the same mass at the GUT scale, $m_{1/2}$.  There is
a GUT scale trilinear coupling among scalars $A_0$.  The last continuous
parameter is taken to be the ratio of the Higgs vacuum expectation values
(VEVs) at the weak scale, $\tan\beta$.  Equivalently, a GUT scale boundary
condition could be applied -- this is the $B$ parameter.  In this framework,
the weak scale Higgs mass parameter $\mu^2$ is derived from the requirement of
electroweak symmetry breaking, thus $\mu$ is derived up to a phase.  To
conserve CP, the phase of $\mu$ is chosen to be $\pm1$, the last parameter of
the model.

Many authors have studied the mSUGRA parameter space from many points of view
\cite{msugradeluge}.  In this paper we explore the full parameter space,
subject to some basic constraints.  The techniques described below allow us to
vary all four mSUGRA parameters freely, while efficiently focusing on the
interesting regions of parameter space.  In contrast, studies in the literature
typically illustrate several 2-parameter slices of the parameter space, usually
holding $A_0$ and $\tan\beta$ fixed.

For each set of parameters, we compute the spectrum using ISAJET \cite{isajet},
and we compute other relevant quantities with DarkSUSY \cite{darksusy}.  The
model confronts current accelerator constraints on sparticle and Higgs masses
\cite{pdg2002}, including preliminary results from LEP concerning a Standard
Model-like Higgs \cite{lephiggs} and the chargino mass \cite{lepsusy}.
Precision results for the $b\rightarrow s\gamma$ branching ratio from the CLEO
and Belle collaborations are applied \cite{bsg} (DarkSUSY computes the NLO
$b\rightarrow s\gamma$).  We mention here that the precise value of the mass of
the top quark $m_t$ can have significant effects on mSUGRA models even within
current error bars.  PDG2002 quotes $m_t=174.3\pm5.1$ GeV \cite{pdg2002}, while
a more recent evaluation from the CDF and D0 collaborations finds
$m_t=178.0\pm4.3$ GeV \cite{topmass}.  In effect we need to take $m_t$ as one
of the model parameters.

The anomalous magnetic moment of the muon $a_\mu=(g_\mu-2)/2$, is sensitive to
new physics such as supersymmetry.  The experimental measurement has improved
greatly in recent years \cite{gminus2}, and there is a hint of a discrepancy
with the Standard Model, though the theoretical calculations of the Standard
Model hadronic contribution are somewhat in doubt
\cite{gminus2theory1,gminus2theory2} due to a discrepancy between evaluations
based on $e^+e^-\rightarrow$ hadrons and those based on hadronic decays of the
$\tau$ lepton.  Quoting the results of Davier et al.\ \cite{gminus2theory1},
the discrepancy is
\begin{eqnarray}
\Delta a_\mu & = & (27\pm10)\times 10^{-10}\; (e^+e^-\;{\rm based}),\\
\Delta a_\mu & = & (12\pm9)\times 10^{-10}\; (\tau\;{\rm based}).
\end{eqnarray}
We apply the constraint that the SUSY contribution must be within $3\sigma$ of
one of these calculations,
\begin{equation}
-15\times10^{-10}<\Delta a_\mu ({\rm SUSY}) <57\times 10^{-10}.
\end{equation}
Furthermore, we will highlight models that fall within $1\sigma$ of either of
these measurements to simulate future constraints,
\begin{equation}
3\times10^{-10}<\Delta a_\mu ({\rm SUSY},1\sigma) <37\times 10^{-10}.
\end{equation}
This latter constraint enforces that the supersymmetric correction be positive,
ruling out $\mu<0$.  Much of the power this constraint has is simply the
requirement of positive $\mu$.

Of astrophysical interest, we compute the relic density of neutralinos in the
model, as well as the neutralino - nucleon elastic scattering cross section.
We are interested in models where the neutralino relic density is consistent
with the cosmological dark matter density, as discussed in the next section.

\section{Dark Matter}

We take the viewpoint that the observed density of cold dark matter is the most
accurate measurement available indicating physics beyond the standard model
\cite{susydmreview}.  The best single measurement to date is that of the
Wilkinson Microwave Anisotropy Probe (WMAP) \cite{wmap}, which measured the
angular power spectrum of the thermal fluctuations in the cosmic microwave
background.  These fluctuations encode information on many cosmological
parameters, including the cosmological densities of many species relative to
the critical density $\Omega_X=\rho_X/\rho_c$, where $\rho_c=3H_0^2/8\pi G$,
and $H_0$ is the Hubble constant.  As is usually done, we take $H_0=100\;h$ km
s$^{-1}$ Mpc$^{-1}$.  The relevant parameters are then the total matter density
$\Omega_M h^2$, the baryon density $\Omega_B h^2$ and the neutrino density
$\Omega_\nu h^2$.  The density of cold dark matter is then just $\Omega_{\rm
CDM}h^2=\Omega_M h^2-\Omega_B h^2-\Omega_\nu h^2$.  The WMAP results are as
follows \cite{wmap}
\begin{eqnarray}
\Omega_M h^2 &=& 0.135^{+0.008}_{-0.009},\\
\Omega_B h^2 &=& 0.0224\pm0.0009,\\
\Omega_\nu h^2 &<& 0.0072 \;(95\% \;{\rm confidence}).
\end{eqnarray}
Other cosmological probes, such as the Sloan Digital Sky Survey (SDSS)
\cite{sdss} give consistent and complementary results.  We naively construct
the cold dark matter density by adding the matter and baryon errors in
quadrature, and simply expanding the lower error bar to accommodate the
possibility of a neutrino component. Thus,
\begin{equation}
\Omega_{\rm CDM} h^2 = 0.1126^{+0.008}_{-0.013}.
\end{equation}
We will consider the following range for the cold dark matter density, allowed
at 95\% confidence,
\begin{equation}
0.087 < \Omega_{\rm CDM} h^2 < 0.129.
\end{equation}

In the next section, we require a likelihood function based on the relic
density of neutralinos.  We will use the WMAP likelihood function, assuming an
asymmetric lognormal distribution.
\begin{eqnarray}
\chi^2&=&\left(\frac{\ln\left(\Omega_{\rm CDM} h^2/0.113\right)}{\sigma}
\right)^2,\\
\sigma &=& 0.068,\; \left(\Omega_{\rm CDM} h^2>0.113\right),\\
\sigma &=& 0.12,\; \left(\Omega_{\rm CDM} h^2<0.113\right).
\end{eqnarray}
The likelihood function is then determined from $\chi^2$ in the usual manner,
$L=\exp(-\chi^2/2)$.

\section{Markov Chain Monte Carlo}

The problem of efficiently scanning high-dimensional parameter spaces appears
in most scientific disciplines.  Clearly, direct grid scans can not be extended
beyond a few dimensions.  Markov Chain Monte Carlo (MCMC) algorithms were
developed to firstly numerically find function minima, and secondly to explore
the region ``near'' the minimum.  The utility of these techniques for data
analysis is obvious: most famously in cosmology the data from the WMAP
satellite were analyzed in this way \cite{wmapmcmc}.  Detailed explanations may
be found in Ref.~\cite{mcmcbook}.

A Markov chain is a sequence of points $P_0 \to P_1 \to P_2 \to \ldots$, with
repetitions allowed, together with a transition probability $W(P_{i+1}|P_i)$
from one point to the next. In a Markov Chain Monte Carlo, a Markov chain is
constructed as follows. The first point of the chain $P_0$ is randomly chosen
according to a prior probability $\Pi(P)$. A new point in the chain is proposed
according to a proposal probability $G(P_{i+1}|P_i)$, and is accepted as part
of the chain according to an acceptance probability $A(P_{i+1}|P_i)$. The
transition probability is then $W(P_{i+1}|P_i) = G(P_{i+1}|P_i)
A(P_{i+1}|P_i)$. It has been proven that if $L(P)$ is a probability
distribution that satisfies the detailed balance condition $W(P_j|P_i) L(P_i) =
W(P_i|P_j) L(P_j)$, then asymptotically, the points $P_i$ in the Markov chain
defined by $W(P_{i+1}|P_i)$ are distributed according to the ``equilibrium''
distribution $L(P)$.

We use the Metropolis algorithm to construct chains of mSUGRA models, with the
hope that the WMAP region will be strongly favored giving scans much more
efficient than grid scans.  A Markov Chain is constructed as follows.  A random
point $P_0$ in parameter space is chosen as the start of the chain, checking to
see that it passes accelerator constraints.  The likelihood function $L\equiv
e^{-\chi^2/2}$ is computed so that $L(P_0)=L_0$.  From this point, a proposal
step is made, to point $P_p$ with likelihood $L_p$.  The Markov chain is
advanced as follows.  If $L_p\ge L_0$ take $P_1=P_p$, i.e., if the new point
has a higher likelihood, take it.  If $L_p<L_0$ take $P_1=P_p$ with probability
$L_p/L_0$, otherwise take $P_1=P_0$, i.e., if the new point has a lower
likelihood, take it with a probability equal to the ratio of likelihoods,
otherwise take the new point equal to the old point.  The process repeats from
$P_1$, and the Markov chain is constructed.  If the proposed point is better,
the chain always advances there.  If the proposed point is worse, the chain
sometimes advances there --- if it is only slightly worse, the chain will
advance there most of the time, if it is much worse, the chain advances only
rarely.  This deceptively simple algorithm is very efficient at traversing a
large-dimensional parameter space, in $O(N)$ time, in contrast to the $O(e^N)$
time for a grid search.

Of course the Metropolis algorithm for advancing the Markov chain is not the
whole story.  Choosing a proposal point based on the $i$-th point in the chain
$P_p(P_i)$ in an optimal way is a subject of much research in the MCMC field.
We now need to know something about the specific problem we are interested in.
We would like to find the region of mSUGRA parameter space that is consistent
with the WMAP data.  This is a strange problem from the point of view of the
usual applications of MCMC, namely we have many more model parameters (four)
than data points (one).  This is not a minimization problem at all, but one of
finding the three-dimensional contours of relic density near the WMAP measured
value.  The degeneracies among parameters are exact.  This contrasts the usual
applications of MCMC, which involve many data points constraining a model with
only a few parameters, with the goal of finding a best fit model and mapping
the likelihood surface around it.

In looking for an optimal strategy in this new context, we try several
strategies for making proposal steps.  The first issue involved is the question
of how big a step should be taken.  We can draw on a simple case for guidance.
Take a likelihood function that is gaussian in the parameters (with zero means
and unit widths) $L=\exp\left(-\sum x_i^2/2\right)$, and take a proposal step
gaussian about the current point, with width $N$ times the width of the
likelihood function $P(p_i)\propto\exp\left(-\sum(p_i-x_i)^2/(2N^2)\right)$.
In the limit of a large number of parameters $D$, the probability to accept the
proposed point is $P({\rm accept})=1-{\rm erf}(N\sqrt{D/8})$.  In this limit
the optimal stepsize (maximizing $N^2P({\rm accept})$, a measure of the
``diffusion velocity'' of the chain) is given by $N=2.381/\sqrt{D}$
\cite{mcmcstep}.  This result assumes that the Markov chain is ``burned in,''
namely it has found the region of reasonable likelihood.  For even a few
parameters these results are acceptable, though for one or two they break down,
e.g.\ for one parameter, it is easy to show that the acceptance probability is
$P({\rm accept}) = (2/\pi) \tan^{-1}(2/N)$ and $N^2 P({\rm accept})$ has no
maximum.\footnote{We have derived an exact expression for P({\rm accept}) as a
function of $N$ and $D$.  In full generality the expression is a double
integral, not obviously tractable for $D>1$.  In the limit of large $D$ the
integrand simplifies, and we reproduce the limit given in Ref.~\cite{mcmcstep}.
For $D\ge2$ the result is
\begin{equation}
P({\rm accept})=\frac{2^{5/2-D}N^{1-D}}{\Gamma\left(\frac{D}{2}\right)
\Gamma\left(\frac{D-1}{2}\right)}
\int_0^\infty dx\,x^{D-1}e^{-x^2/2}\int_0^x dy\,y^{D-2} e^{-y^2/2N^2}
\sum_\pm{\rm erf}\left(\frac{\sqrt{x^2-y^2}\pm x}{N\sqrt{2}}\right).
\end{equation}
For $D=2$ we find again that the function $N^2P({\rm accept})$ has no maximum,
in fact for $N\gg1$, $P({\rm accept})\propto N^{-D}$.}
However, it has been shown that $N\approx2.4/\sqrt{D}$ is a reasonable
approximation even for $D=1,2$ \cite{mcmccosmology}.  What we can draw from
this is an optimal acceptance probability, $P({\rm accept})=0.2338$.  This
means that the fastest exploration of the parameter space happens when only
about 1/4 of the proposals are accepted.  This allows the stepsize to remain
relatively large, but still a reasonable fraction of points are accepted.
Based on this result, we apply a simple method for adapting the stepsize.  If
too many proposals are accepted in a row, it can be inferred that the
probability to step is too large to optimally sample the parameter space.  The
indication is that the stepsize is too small and should be increased.
Likewise, if too many proposals are rejected in a row, the acceptance
probability may be too small and the stepsize may be too large.  For example,
based on an acceptance probability of 1/4, three consecutive acceptances is
already unlikely at the 98.4\% level, and the stepsize should be increased.
Aiming for a similar probability for the other extreme, if 14 proposals are
rejected in a row, the stepsize should be decreased.  In both cases we choose
to change the stepsize by a factor of 2.  We have made crude explorations of
these thresholds, and found that changing the stepsize for 3--4 consecutive
acceptances or 6--8 consecutive rejections works fairly well.  The exact
numbers can be adjusted, but we find that the threshold for rejections should
be higher than for acceptance.  Equivalently, an acceptance probability less
than 0.5 is ideal.  Setting the threshold for rejections at twice the threshold
for acceptances corresponds to an acceptance probability of 38\%.

The second issue we discuss is that of step direction.  We know that there are
exact degeneracies, as we place only one constraint on many parameters.  We
would like to step in a direction that is likely to find another good
point. When the chain point is close to the WMAP value of $\Omega_{\rm CDM}
h^2$, we would like the next point to lie (approximately) on the $\Omega_{\rm
CDM} = \Omega_{\rm CDM,WMAP}$ surface.  The most obvious thing to do is to
calculate the likelihood gradient at the current point in the Markov chain, and
step orthogonally to the gradient.  We do this when 4 (or the number of
dimensions) consecutive rejections are collected, as soon as the current
likelihood is reasonably good (within 3--4$\sigma$).  With 4 directions along
which the difference in likelihood is known, the gradient vector can be
computed.\footnote{We compute the gradient of $\Omega_{\rm CDM}h^2$ rather than
the gradient of the likelihood function. The two gradients share the same
direction since our likelihood is a function of $\Omega_{\rm CDM}h^2$ only.
However, the gradient of the likelihood function vanishes on the WMAP surface,
thus we use the former gradient.}
After this, steps are taken in arbitrary directions perpendicular to the
gradient vector.  A refinement of this procedure is to allow a small step along
the gradient, though much smaller than that perpendicular.  This can be done in
a fixed manner (e.g. 10:1 ratio, 20:1 ratio), or adaptively.  We employ the
gradient technique for most of the mSUGRA scans discussed in this paper.
We remark here that this choice of proposal probability depends through the
gradient and stepsize on more than just the previous point in the Markov
chain. As a consequence, its associated transition probability does not satisfy
the condition of detailed balance, and we cannot immediately prove that our
gradient algorithm converges to the likelihood distribution $L(P)$.
Nevertheless, based on the intuition that an ``equilibrium'' distribution
should not depend on the initial conditions or on the way equilibrium is
approached, we expect that the points in the Markov chain are distributed
according to the likelihood $L(P)$ after the chain has ``converged.'' It must
however be kept in mind that the final distribution of points in parameter
space depends also on where we choose the initial point of each chain, i.e.
our prior distribution.

Another method for using directional information involves the interactions
among several Markov chains.  The basic assumption is that if two Markov chains
have found good regions, then the line connecting their current positions might
also be in the good region.  For each chain, a random second chain is selected,
and the step taken is in the direction (positive or negative) of this chain.
This is the ``snooker'' algorithm.  A few of our scans use this algorithm,
trying interactions between 8 chains, or 128 chains.  We find neither does a
great job of covering the full surface, and that perhaps more chains are
required.  This difficulty is possibly related to the fact that we are trying
to trace a surface instead of finding a global minimum.

\section{Scanning the mSUGRA Parameter Space}

We now explore the mSUGRA parameter space using the techniques of the previous
section.  First, we employ a very crude grid search, taking only a few points
along the $\tan\beta$ and $A_0$ axes.  Roughly 1\% of the scanned points pass
the WMAP $2\sigma$ relic density cut.  Crudely, this is the 4-dimensional
``efficiency'' of the grid search.

As the first illustration of the MCMC method, we attempt to duplicate a typical
mSUGRA scan from the literature.  We choose Fig.~3 of Ref.\cite{baerpaper},
which illustrates the usual basic features -- at moderate $\tan\beta$,
acceptable relic density occurs only for the stau coannihilation region at
small $m_0$ or the focus point region at large $m_0$ where chargino
coannihilations are important.  This model has $A_0=0$, $\tan\beta=30$,
$\mu>0$, and $m_t=175$ GeV.  Fixing these parameters, we make a 2 dimensional
scan in $m_0$ and $m_{1/2}$.  In Fig.~\ref{fig:points} we plot every point
where the relic density was calculated.  Points that pass the WMAP relic
density cut are highlighted in red, and those with relic density below the WMAP
region are highlighted in blue.  It is clear that the MCMC technique does an
excellent job of automatically finding the WMAP preferred region.  For these
scans, the gradient method was used.  The usual features are clear in that the
acceptable regions are the stau coannihilation funnel at small $m_0$ and the
focus point region at large $m_0$. About 40,000 points have been generated in
this scan. Notice that only a small fraction of points lies outside the good
WMAP region, so few that they can be counted individually on the figure. This
indicates that the Markov chains converge to the good region rapidly.

We now extend the scans to the full 4 dimensional parameter space.  We are
searching for the 3 dimensional surface near which the relic density is
acceptable.  We separately scan both signs of $\mu$, and take three values for
the top quark mass: $m_t=174.3$ GeV (PDG 2002 central value \cite{pdg2002}),
$m_t=178.0$ GeV (current CDF/D0 central value \cite{topmass}), and $m_t=181.7$
GeV (within $1\sigma$ of the central value).  Using the MCMC techniques of the
previous section, predominantly the gradient method, but also with some scans
using the snooker algorithm, we find efficiencies of 20\%-30\% to find points
that pass the relic density cut.  This is remarkable in that the acceptable
region is found automatically.  In total, we have sampled some 2.4 million
models, with more than 500 thousand within the WMAP region.  The full extent of
the scans is as follows,
\begin{eqnarray}
50\;{\rm GeV} & < m_0 < & 50\;{\rm TeV},\\
50\;{\rm GeV} & < m_{1/2} < & 20\;{\rm TeV},\\
-20\;{\rm TeV} & < A_0 < & 20\;{\rm TeV},\\
2 & < \tan\beta < & 60.
\end{eqnarray}
MCMC techniques allow us to scan freely in all 4 parameters.  Since the MCMC
algorithm is essentially linear in the number of parameters, the utility of the
technique for more complex models is clear.

As we are scanning freely in $A_0$, we must be careful about unacceptable
vacua, both potential directions unbounded from below (UFB) and the so-called
charge and/or color breaking (CCB) vacua that occur if e.g.\ the stop gets a
VEV \cite{ufb_and_ccb}.  We apply a simple constraint \cite{ccbconstraint} that
(conservatively) allows for CCB minima as long as they are long-lived,
\begin{equation}
A_t^2+3\mu^2<7.5\left(m^2_{\tilde{t}_L}+m^2_{\tilde{t}_R}\right),
\end{equation}
and the parameters are taken at the weak scale.  This constraint in particular
excludes the stop from getting a VEV, removing a small number of models where
$|A_0|\gg m_0$.  Even this constraint may not be permissive enough; recent work
indicates that radiative corrections soften some of these constraints
considerably \cite{ccbnewconstraint}.

\section{Projections of mSUGRA Parameter Space}

After performing these scans, we have a 4 dimensional cloud of models, a
significant fraction of which pass the WMAP relic density cut.  In order to
display it easily, we project onto the six two-dimensional coordinate planes.
We will illustrate both theoretical (involving the 4 mSUGRA parameters) and
phenomenological (e.g.\ LSP mass and cross section) examples.

As an application to dark matter searches, we consider the spin-independent
neutralino-nucleon elastic scattering cross section, $\sigma_{\rm SI}$.  We
plot the neutralino mass $m_\chi$ vs.\ $\sigma_{\rm SI}$ in
Fig.~\ref{fig:sigsi}, for every model passing the WMAP constraint.  We
illustrate the DAMA preferred region \cite{dama}, current experimental bounds
from EDELWEISS \cite{edelweiss} and CDMS \cite{cdmscurrent}, and the future
reach of several experiments, running (CDMS II \cite{cdmsII}, CRESST II
\cite{cresstII}) and proposed (GENIUS \cite{genius}, CryoArray \cite{cryoarray}
and XENON \cite{xenon}).  For an extensive collection of data and projections
for dark matter experiments see Ref.~\cite{dmtools}.  We have highlighted
models having a $\Delta a_\mu({\rm SUSY})$ within $1\sigma$ of the current
experimental bounds to illustrate the correlation between $a_\mu$ and
$\sigma_{\rm SI}$ \cite{gm2ddcorr}.

In Fig.~\ref{fig:projwmap} and Fig.~\ref{fig:projnowmap}, we plot all six
2-dimensional projections of the 4 dimensional mSUGRA parameter space, with one
subplot each for $\mu=\pm1$.  There is one plot for every parameter pair in
$m_0$, $m_{1/2}$, $A_0$ and $\tan\beta$.  Fig.~\ref{fig:projwmap} illustrates
only those models that obey the WMAP constraint, while
Fig.~\ref{fig:projnowmap} illustrates all models passing the accelerator
constraints.  Seen in projection, the WMAP constraint does not look very
powerful, but it typically singles out very thin surfaces as seen in
Fig.~\ref{fig:points}.  These thin surfaces disappear in projection, since
their exact position in a 2-dimensional projection is a function of the other
two parameters.

\section{Discussion}

We have shown that MCMC is a powerful technique for scanning high dimensional
parameter spaces in supersymmetry by demonstrating that it can navigate the
mSUGRA parameter space and find regions with relic density acceptable to WMAP.
This is a heartening conclusion.  On the face of it, finding acceptable mSUGRA
models is a very difficult problem for any function minimization technique.
There are several issues involved.  First, there are exact degeneracies.  What
we really need is a contouring algorithm.  The gradient method alleviates this
difficulty.  Second, the acceptable regions tend to be on the edges of the
allowed parameter space.  Third, there are acceptable regions that are
disjoint, or nearly so.  A certain amount of brute force is required; what MCMC
allows for is a considerable reduction in the amount of brute force necessary.

A more correct approach to the parameter scanning would use likelihoods based
on accelerator data rather than hard constraints.  In particular, the
$b\rightarrow s\gamma$ branching ratio and $\Delta a_\mu({\rm SUSY})$ would be
easy to implement in this way.  Limits on particle masses could be expressed as
likelihoods as well, though most values would be equally likely, with the
likelihood dropping as the ``limit'' was approached.  In fact, an analysis of
the $\chi^2$ fit for the combination $b\rightarrow s\gamma$, $\Delta a_\mu({\rm
SUSY})$ and $\Omega_{\rm CDM}h^2$ has been performed \cite{baerchi2}.  An
obvious extension of our work would be to use this likelihood function with
three data points to do an MCMC scan of parameter space as we have done.
However, none of the current accelerator constraints is nearly as powerful as
the relic density constraint.  From the viewpoint of finding acceptable models,
we have used the most important constraint in the MCMC scans.

The likelihood function applied need not be related to real data.  To scan an
interesting region in parameter space, a likelihood function could be designed
to favor that region.  For example, in addition to including the WMAP relic
density measurement, a likelihood function could be constructed to favor large
elastic scattering cross sections, or any other interesting signal.  In this
way, models consistent with the WMAP relic density AND with high cross sections
would be preferentially chosen.  We leave explorations of this possibility to
future work.  Optimistically, if supersymmetry is discovered in e.g.\ LHC data,
new terms in the likelihood function would be required.  The gradient method
can be adapted when the number of terms is less than the number of parameters:
the gradient of each constraint is computed in turn, and steps are taken in the
parameter subspace orthogonal to all such gradients.

It is tempting to ascribe a statistical significance to the density of models
calculated (more correctly, using the points of the Markov Chain which excludes
rejected points and possibly includes accepted points multiple times).  We urge
caution here, simply stating that the higher density regions tend to be
associated with acceptable relic density and as such are more likely.

In a sense (not terribly well-defined), mSUGRA is disfavored by the WMAP
results in that getting a reasonable relic density is ``difficult,'' usually
requiring coannihilations or strong resonance effects.  We plan to explore more
general model frameworks which may relax these difficulties.  MCMC will allow
such studies, while not requiring particularly immense computational resources.

\FIGURE{
\epsfig{file=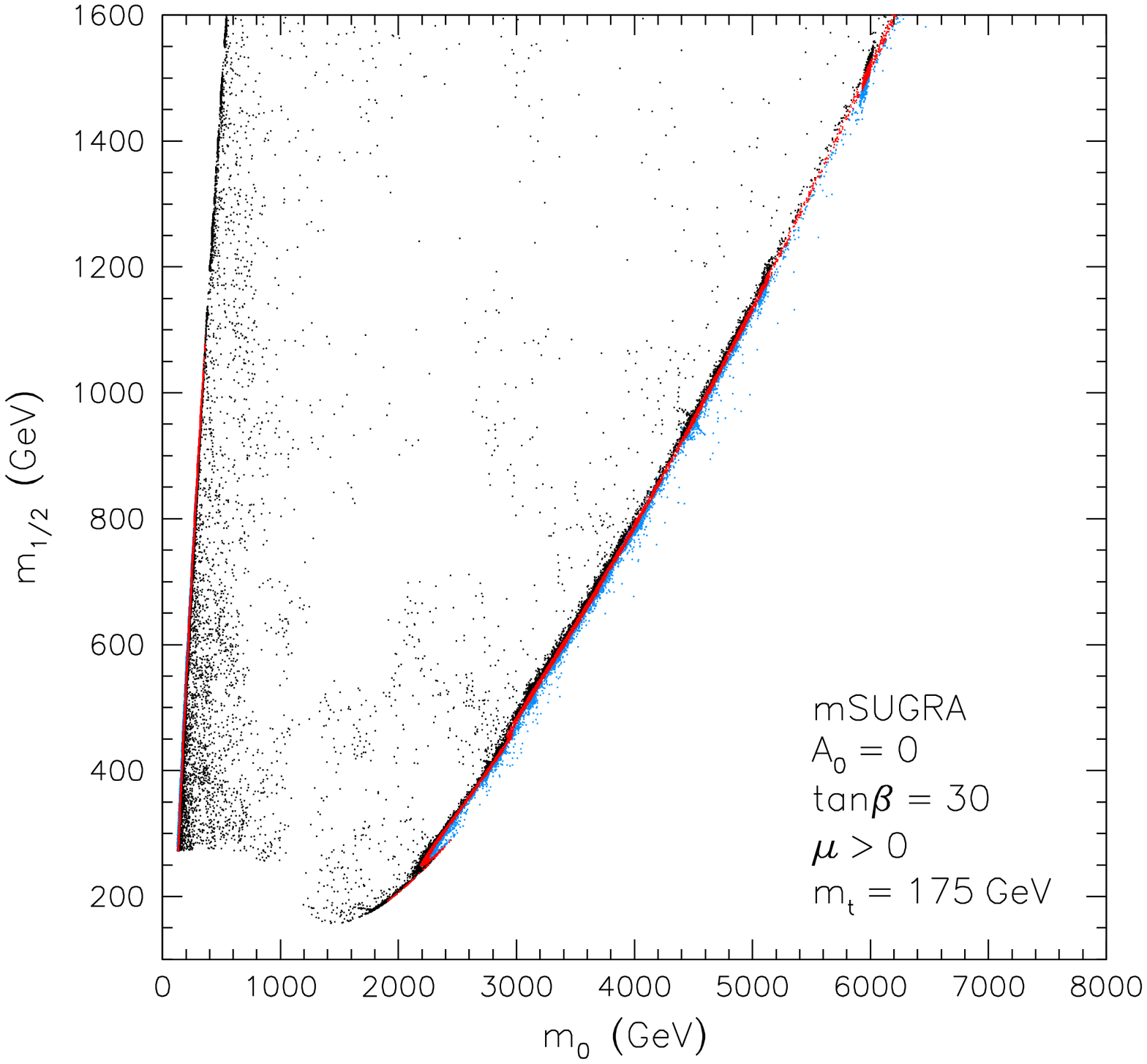,width=\textwidth}
\caption{MCMC scan in $m_0$ and $m_{1/2}$, holding other parameters fixed.  All
  points where the relic density was calculated are included.  Compare with
  Fig.~3 of Ref.~\cite{baerpaper}.  The strong preference for ``interesting''
  regions is clear.  Black points have $\Omega_\chi h^2$ more than $2\sigma$
  too large according to the WMAP constraint, red points are within the
  $2\sigma$ region, and blue points have too little relic density at the
  2$\sigma$ level -- perfectly acceptable, but not as the sole constituent of
  dark matter.  The gradient method was used for these scans.  For some chains,
  the gradient estimate was not very good, e.g.\ the feature at $m_0=6$ TeV,
  $m_{1/2}=1.5$ TeV.  These runs allow steps along the gradient of 5\% or 10\%
  of the step length.  This plot is meant to illustrate the fact that very
  little time is spent exploring ``uninteresting'' regions of the parameter
  space.  About 40,000 models are shown here, equivalent to a 200x200 grid
  search.  For two dimensions, MCMC may not be a big improvement, but in more
  than two dimensions, it gains a clear advantage.  Note that since we include
  all points, these are NOT the Markov Chains.  The Markov Chains exclude
  rejected points and include some points multiple times.  The Markov chains do
  not however reflect the computational efficiency that is a major concern in
  this work.}
\label{fig:points}
}

\FIGURE{
\epsfig{file=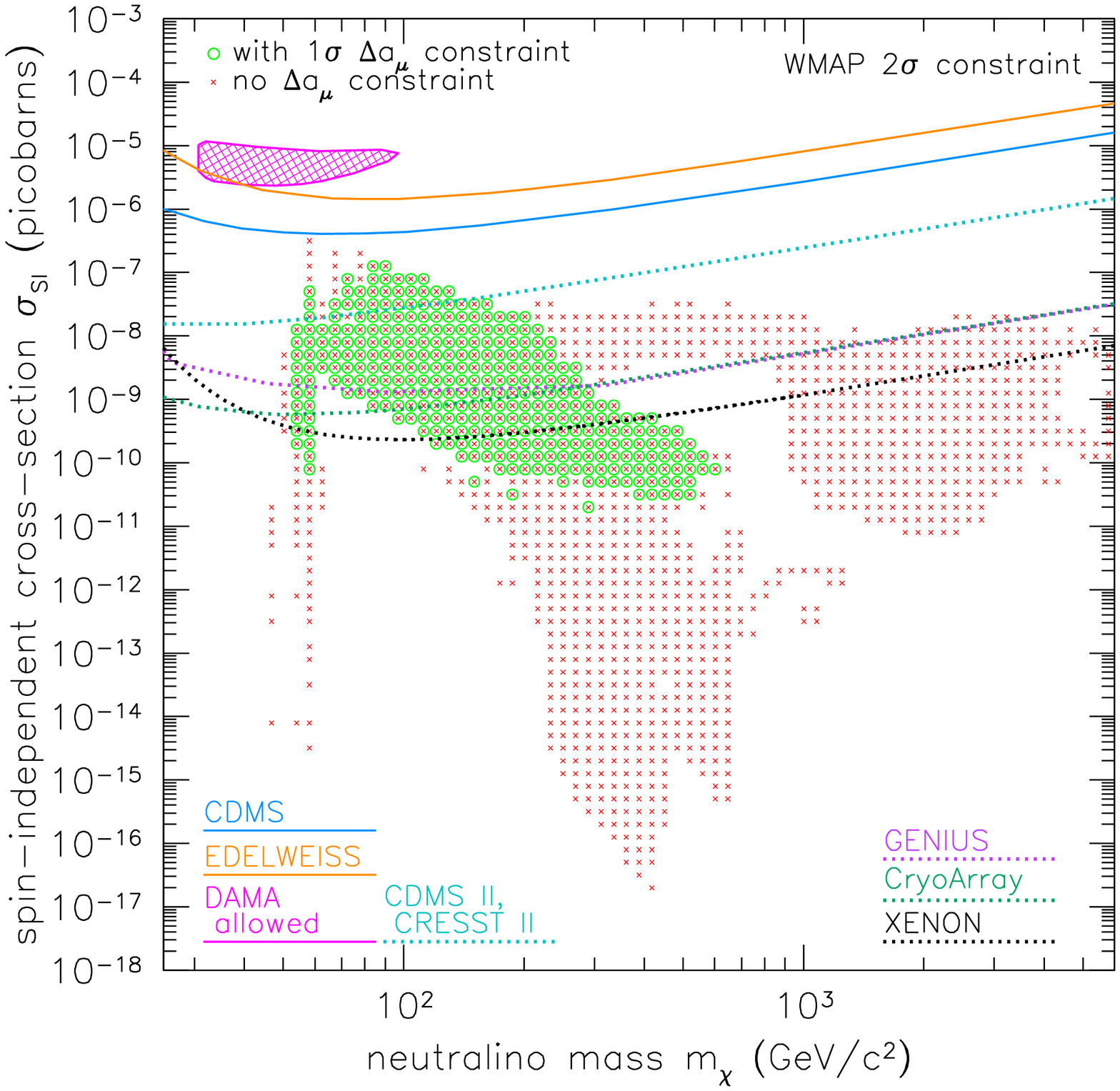,width=\textwidth}
\caption{Spin-independent elastic scattering cross section of neutralinos on
  protons.  The models have been collected in bins in a square grid.  Current
  experimental limits from EDELWEISS and CDMS II are shown, as well as the
  region favored by the DAMA annual modulation result.  The proposed reach of
  the CDMS II, CRESST II, CryoArray, GENIUS, and XENON experiments is also
  shown.  Models falling withing 1$\sigma$ of the current measurement of
  $a_\mu$ are highlighted by green circles, though not all models in those bins
  necessarily pass the $a_\mu$ cut.}
\label{fig:sigsi}
}

\FIGURE{
\epsfig{file=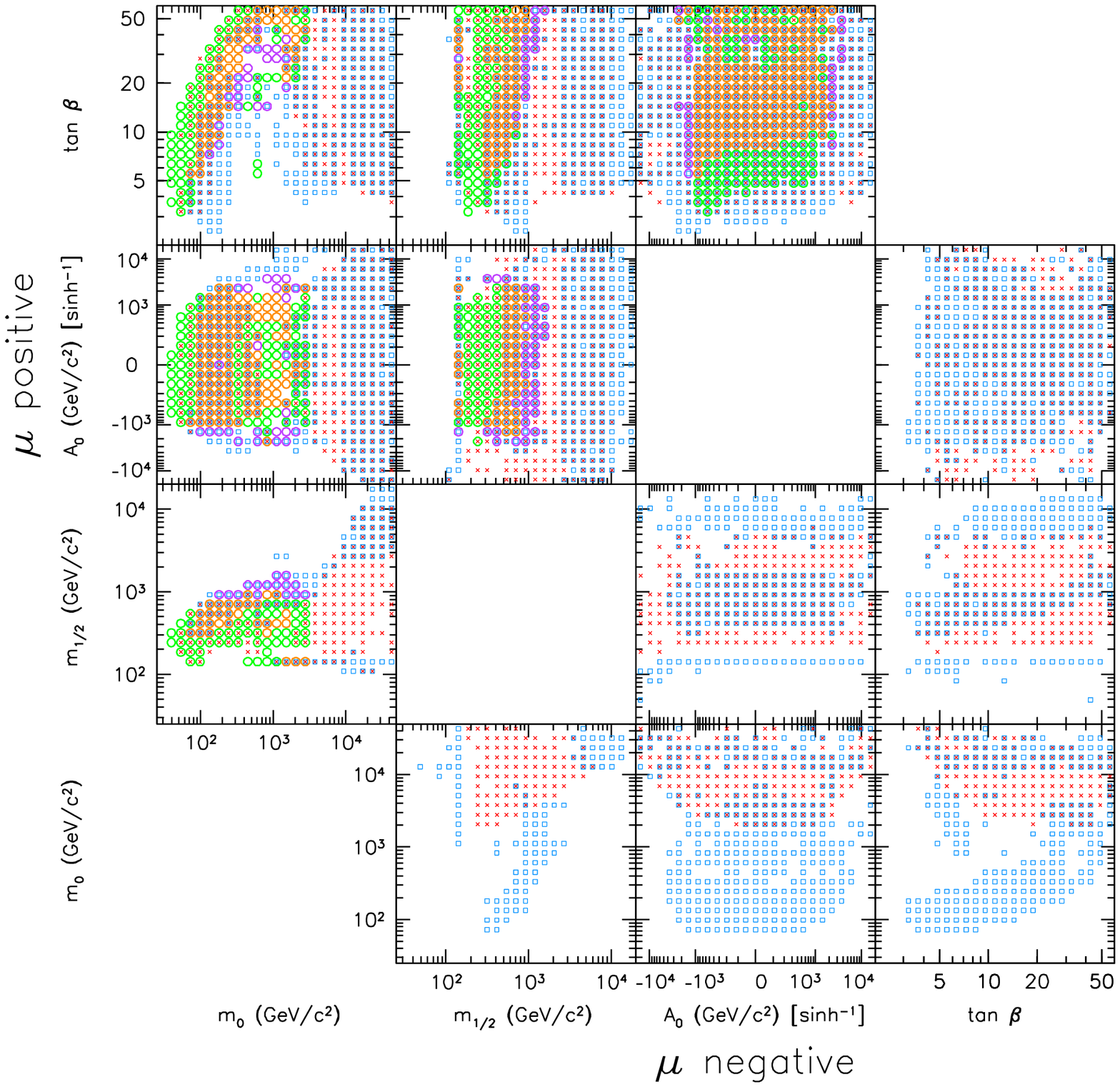,width=\textwidth}
\caption{Projections of mSUGRA parameter space obeying the WMAP constraint on
  relic density.  All six 2-dimensional projections of the four continuous
  parameters $m_0$, $m_{1/2}$, $A_0$ and $\tan\beta$ are illustrated, with
  $\mu$ positive in the upper left and $\mu$ negative in the lower right.  The
  green, orange and purple circles indicate models within 1$\sigma$ of the
  current measurement of the muon $a_\mu$.  Green indicates that all models
  passing the $a_\mu$ cut will be detectable by the proposed XENON experiment,
  orange indicates that some such models will be detectable, and purple
  indicates that no such models will be detectable.  The red crosses show
  models that fail the $a_\mu$ constraint, and will be detectable by XENON, and
  the blue squares show models that fail the $a_\mu$ constraint, and will not
  be detectable.  In projection, the WMAP constraint does not seem to be very
  powerful, in contrast to Fig.~\ref{fig:points}.  This is simply because the
  thin sheet of allowed parameter space is not perpendicular to any particular
  one of the mSUGRA parameters.  Note that $m_0$, $m_{1/2}$, and $\tan\beta$
  are plotted logarithmically, while $A_0$ is plotted as
  $\sinh^{-1}(A_0/100\;{\rm GeV})$, with the first non-zero tickmark being 100
  GeV.}
\label{fig:projwmap}
}

\FIGURE{
\epsfig{file=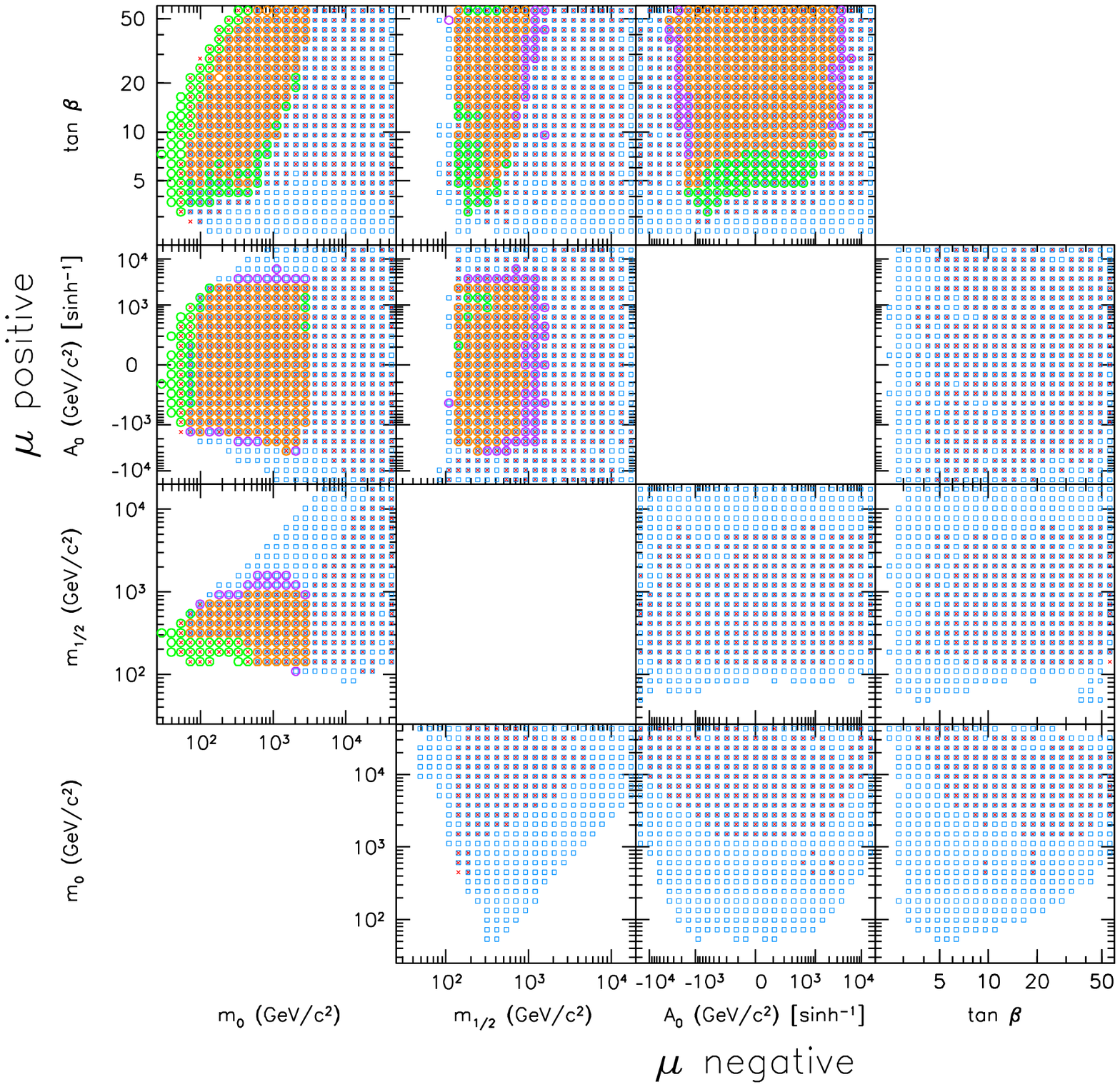,width=\textwidth}
\caption{Projections of mSUGRA parameter space without the WMAP constraint on
  relic density.  The points are plotted as in Fig.~\ref{fig:projwmap}.  The
  WMAP constraint does not appear to be making a huge difference when viewed in
  projection, though in fact the 4-dimensional volume occupied by acceptable
  models is quite small.}
\label{fig:projnowmap}
}

\acknowledgments E.~A.~B. thanks Phil Marshall and John Peterson for selflessly
educating him on MCMC techniques, and thanks Michael Peskin for many useful
conversations on mSUGRA models.  We thank the Michigan Center for Theoretical
Physics for hospitality during the Dark Side of the Universe workshop, where
some of this research was done.  This work was supported in part by the
U.S. Department of Energy under contract number DE-AC03-76SF00515.

\end{document}